\documentclass[twocolumn,showpacs,preprintnumbers,amsmath,amssymb]{revtex4}
\usepackage{graphicx}
\usepackage{dcolumn}
\usepackage{bm}

\begin{document}

\preprint{APS/123-QED}

\title{Larmor times for transmission and reflection}

\author{N. L. Chuprikov}
\affiliation{%
Tomsk State Pedagogical University, 634041, Tomsk, Russia}
\altaffiliation[Also at
]{Physics Department, Tomsk State University.}

\date{\today}

\begin{abstract}

On the basis of quant-ph/0405028 we define the Larmor times for transmission and
reflection. These times are valid both for the stationary and time-dependent
scattering processes, without any restrictions on the shape of Gaussian-like wave
packets. We show that in the stationary case both the Larmor times coincide with the
corresponding dwell times obtained in quant-ph/0502073. The Larmor-time concept
gives the way to verify the approach presented in quant-ph/0405028.

\end{abstract}
\pacs{03.65.Ca, 03.65.Xp }
\maketitle
\newcommand{\Api}{A_{in}}
\newcommand{\Ami}{B_{in}}
\newcommand{\Apo}{A_{out}}
\newcommand{\Amo}{B_{out}}
\newcommand{\bpi}{a_{in}}
\newcommand{\bmi}{b_{in}}
\newcommand{\bpo}{a_{out}}
\newcommand{\bmo}{b_{out}}
\newcommand{\api}{a_{in}}
\newcommand{\ami}{b_{in}}
\newcommand{\apo}{a_{out}}
\newcommand{\amo}{b_{out}}
\newcommand{\ppp}{\mbox{\hspace{5mm}}}
\newcommand{\ooo}{\mbox{\hspace{3mm}}}
\newcommand{\ooa}{\mbox{\hspace{1mm}}}
\section{Introduction}

By our approach (see \cite{Ch1}) the state of the quantum ensemble of identically
prepared particles tunneling through a symmetrical time-independent potential
barrier represents a coherent superposition of two states: that of particles which
are finally transmitted through the barrier, and that of particles which are finally
reflected by it. This is valid for any stage of this scattering process. For a given
symmetrical potential barrier and initial state of the whole ensemble of particles
there is an unique pair of solutions to the Schr\"odinger equation, which describe
the coherently evolved alternative sub-processes, transmission and reflection.

Note, the fact that the above decomposition is valid, in particular, for the first
stage of scattering does not at all mean that the future of a starting particle is
predetermined in our approach. Of course, this is not the case. The knowledge of the
wave functions for transmission and reflection does not permit us to predict the
behavior of one particle in one experiment. As regards the subensembles of
transmitted and reflected particles, their history has a deterministic character
and, thereby, is predictable. This agrees entirely with the basic principles of
quantum mechanics.

Our approach opens new possibilities in solving the so-called tunneling time problem
to have been of great controversy for the last decades. In \cite{Ch1} we have
introduced the so-called (exact and asymptotic) group tunneling times for
transmission and reflection, and in \cite{Ch2} we have defined the dwell times for
these two sub-processes. These time concepts supplement each other. The group
tunneling times for transmission an reflection are formally valid in the general
case. However, in fact they more fit only for timing a particle in a well-localized
state. As regards sufficiently wide (in $x$-space) wave packets, the corresponding
expectation value of the particle's position is badly defined. Moreover, in this
case, reflected particles do not enter, on the average, into the barrier region.
Thus, the exact group reflection time is zero in this case, what is evident to be a
very rough description of the reflection process.

In the case of infinitely wide wave packets, the time spent by a particle in the
barrier region is described by the dwell time. However, this time concept, too, has
shortcomings: 1) it has been introduced intuitively, rather than from the first
principles; 2) and, besides, it is valid only for the stationary scattering problem.

In addition to the above, there is once more shortcoming to be common both for the
group and dwell time concepts: they do not give the way of measuring the above
tunneling times for transmission and reflection. The first concept does not give the
way of measuring the (average) time of entering to-be-transmitted (or
to-be-reflected) particles in the barrier region. Similarly, the second one does not
suggest the way of how experimentally to extract the information about the
individual properties of the subensembles of to-be-transmitted or to-be-reflected
particles in the barrier region.

It is evident that to distinguish to-be-transmitted and to-be-reflected particles is
impossible. In order to measure the (average) times spent by the subensembles of
transmitted and reflected particles in the barrier region, we have to place
measuring detectors far from the barrier region. Perhaps, the most suitable physical
effect which can be assumed as a basis for such measurements is the Larmor
precession of the particle's spin, in a weak magnetic field confined to the barrier
region.

As is known, the idea to use the Larmor precession as clocks was proposed by Baz'
\cite{Baz} and developed later by Rybachenko \cite{Ryb} and B\"{u}ttiker \cite{But}
(see also \cite{Lea,Kob}). Our aim is to define the Larmor times for transmission
and reflection, on the basis of the wave functions found in \cite{Ch1}. Besides, we
consider, of great importance for understanding the tunneling phenomenon is to
answer the questions posed in \cite{Lea}. One of them concerns the appearance of a
superluminal transmission of a particle through sufficiently opaque and wide
potential barriers (the Hartman effect). Another question relates to the influence
of the magnetic field confined to a spatial region behind the barrier, on the spin
of reflected particles.

Apart from these two question one has to explain the effect found by B\"{u}ttiker
(see \cite{But}). He showed that the magnetic field aligns the particle's spin with
the field. This result is evident to be paradoxical too. For the operator of the
spin $z$-component commutes with the Hamiltonian in this problem. The expectation
value of this operator should be constant individually both for transmitted and
reflected particles.

We have to stress once more (see \cite{Ch1}) that averaging any observable (and the
well-known Born's formula for the expectation values) has a physical sense only for
the so-called "elementary" states, which cannot be decomposed onto macroscopically
distinguishable ones. The notion of the integrals of motion is valid, too, only for
"elementary" states. In the case considered the states of the subensembles of
transmitted and reflected particles are just the states of such kind. As regards
averaging over the {combined} state of the whole ensemble of tunneling particles,
this procedure is meaningless for it. Just calculating the "expectation values" of
physical quantities, for "combined" states, leads to paradoxes.

\section{Tunneling a spinning particle through a symmetrical finite potential
barrier: Setting the problem}

Let us consider the quantum ensemble of electrons moving along the $x$-axis and
interacting with the symmetrical time-independent potential barrier $V(x)$ and small
magnetic field (parallel to the $z$-axis) confined to the finite spatial interval
$[a,b]$ $(a>0)$; $V(x-x_c)=V(x_c-x)$; $x_c=(b+a)/2$; $d=b-a$ is the barrier width.
Let this ensemble be a mixture of two parts. One of them consists from electrons
with spin parallel to the magnetic field. Another is formed from particles with spin
down.

Let at $t=0$ the state of this mixture is described by the spinor
\begin{eqnarray} \label{1}
\Psi_{full}^{(0)}(x)=\frac{1}{\sqrt{2}}\left(\begin{array}{c} 1 \\ 1
\end{array} \right)\psi_{full}^{(0)}(x),
\end{eqnarray}
where $\psi_{full}^{(0)}(x)$ is a normalized function. Without loss of generality we
will suppose that
\[<\psi_{full}^{(0)}|\hat{x}|\psi_{full}^{(0)}>=0,\ooa
<\psi_{full}^{(0)}|\hat{p} |\psi_{full}^{(0)}> =\hbar k_0 > 0,\]
\[<\psi_{full}^{(0)}|\hat{x}^2|\psi_{full}^{(0)}> =l_0^2,\]
here $l_0$ is the wave-packet's half-width at $t=0$ ($l_0<<a$); $\hat{x}$ and
$\hat{p}$ are the operators of the particle's position and momentum, respectively.

So, we will consider the case, when the initial spin coherent state (\ref{1}) is the
engine state of $\sigma_x$ with the eigenvalue 1 (the average spin of the ensemble
of incident particles is oriented along the $x$-direction); hereinafter, $\sigma_x,$
$\sigma_y$ and $\sigma_z$ are the Pauli spin matrices.

For electrons with spin up (down), the potential barrier effectively decreases
(increases), in height, by the value $\hbar\omega_L/2$; here $\omega_L$ is the
frequency of the Larmor precession; $\omega_L=2\mu B/\hbar,$ $\mu$ denotes the
magnetic moment. The corresponding Hamiltonian has the following form,
\begin{eqnarray} \label{200}
\hat{H}=\frac{\hat{p}^2}{2m}+V(x)-\frac{\hbar\omega_L}{2}\sigma_z, \ooo if\ooo
x\in[a,b];\nonumber\\ \hat{H}=\frac{\hat{p}^2}{2m}, \ooo otherwise;
\end{eqnarray}
here $m$ is the mass of a particle. For $t>0$, due to the influence of the magnetic
field, the states of particles with spin up and down become different. The
probability to pass the barrier is different for them. Let for any value of $t$ the
spinor to describe the state of particles read as
\begin{eqnarray} \label{2}
\Psi_{full}(x,t)=\frac{1}{\sqrt{2}}\left(\begin{array}{c} \psi_{full}^{(\uparrow)}(x,t) \\
\psi_{full}^{(\downarrow)}(x,t) \end{array} \right).
\end{eqnarray}

In accordance with \cite{Ch1}, each of these two spinor components can be uniquely
presented as a coherent superposition of two probability fields to describe
transmission and reflection:
\begin{eqnarray} \label{3}
\psi_{full}^{(\uparrow)}(x,t)=
\psi_{tr}^{(\uparrow)}(x,t)+\psi_{ref}^{(\uparrow)}(x,t);
\nonumber\\\psi_{full}^{(\downarrow)}(x,t)=
\psi_{tr}^{(\downarrow)}(x,t)+\psi_{ref}^{(\downarrow)}(x,t).
\end{eqnarray}
(remind (see \cite{Ch1}) that $\psi_{ref}(x,t)\equiv 0$ for $x\ge x_c$). As a
consequence, the same decomposition takes place for spinor (\ref{2}):
\begin{eqnarray} \label{4}
\Psi_{full}(x,t)= \Psi_{tr}(x,t)+\Psi_{ref}(x,t).
\end{eqnarray}
We will suppose that all the wave functions for transmission and reflection are
known. It is important to stress here (see \cite{Ch1}) that
\begin{eqnarray} \label{100}
<\psi_{full}^{(\uparrow\downarrow)}(x,t)|\psi_{full}^{(\uparrow\downarrow)}(x,t)>
=T^{(\uparrow\downarrow)}+R^{(\uparrow\downarrow)}=1
\end{eqnarray}
where
\begin{eqnarray} \label{101}
T^{(\uparrow\downarrow)}=<\psi_{tr}^{(\uparrow\downarrow)}(x,t)|
\psi_{tr}^{(\uparrow\downarrow)}(x,t)>=const;
\end{eqnarray}
\begin{eqnarray} \label{102}
R^{(\uparrow\downarrow)}=<\psi_{ref}^{(\uparrow\downarrow)}(x,t)|
\psi_{ref}^{(\uparrow\downarrow)}(x,t)>=const;
\end{eqnarray}
$T^{(\uparrow\downarrow)}$ and $R^{(\uparrow\downarrow)}$ are the (real)
transmission and reflection coefficients, respectively, for particles with spin up
$(\uparrow)$ and down $(\downarrow)$. Let further
$T=(T^{(\uparrow)}+T^{(\downarrow)})/2$ and $R=(R^{(\uparrow)}+R^{(\downarrow)})/2$
be quantities to describe the whole ensemble of particles.

\section{Time evolution of the spin polarization of tunneling particles}

To study the time evolution of the average particle's spin, we have to find the
expectation values of the spin projections $\hat{S}_x$, $\hat{S}_y$ and $\hat{S}_z$.
Note, for any $t$
\begin{eqnarray} \label{5}
<\hat{S}_x>_{full}\equiv
\frac{\hbar}{2}\sin(\theta_{full})\cos(\phi_{full})\nonumber\\=\hbar \cdot
\Re(<\psi_{full}^{(\uparrow)}|\psi_{full}^{(\downarrow)}>);
\end{eqnarray}
\begin{eqnarray} \label{6}
<\hat{S}_y>_{full}\equiv
\frac{\hbar}{2}\sin(\theta_{full})\sin(\phi_{full})\nonumber\\=\hbar \cdot
\Im(<\psi_{full}^{(\uparrow)}|\psi_{full}^{(\downarrow)}>);
\end{eqnarray}
\begin{eqnarray} \label{7}
<\hat{S}_z>_{full}\equiv
\frac{\hbar}{2}\cos(\theta_{full})\nonumber\\=\frac{\hbar}{2}
\left[<\psi_{full}^{(\uparrow)}|\psi_{full}^{(\uparrow)}>
-<\psi_{full}^{(\downarrow)}|\psi_{full}^{(\downarrow)}>\right].
\end{eqnarray}
Similar expressions are valid for transmission and reflection:
\begin{eqnarray} \label{8}
<\hat{S}_x>_{tr}=\frac{\hbar}{T}
\Re(<\psi_{tr}^{(\uparrow)}|\psi_{tr}^{(\downarrow)}>);
\end{eqnarray}
\begin{eqnarray} \label{9}
<\hat{S}_y>_{tr}=\frac{\hbar}{T}
\Im(<\psi_{tr}^{(\uparrow)}|\psi_{tr}^{(\downarrow)}>);
\end{eqnarray}
\begin{eqnarray} \label{10}
<\hat{S}_z>_{tr}=\frac{\hbar}{2T}
\left(<\psi_{tr}^{(\uparrow)}|\psi_{tr}^{(\uparrow)}>
-<\psi_{tr}^{(\downarrow)}|\psi_{tr}^{(\downarrow)}>\right);
\end{eqnarray}
\begin{eqnarray} \label{11}
<\hat{S}_x>_{ref}=\frac{\hbar}{R}
\Re(<\psi_{ref}^{(\uparrow)}|\psi_{ref}^{(\downarrow)}>);
\end{eqnarray}
\begin{eqnarray} \label{12}
<\hat{S}_y>_{ref}=\frac{\hbar}{R}
\Im(<\psi_{ref}^{(\uparrow)}|\psi_{ref}^{(\downarrow)}>);
\end{eqnarray}
\begin{eqnarray} \label{13}
<\hat{S}_z>_{ref}=\frac{\hbar}{2R}
\left(<\psi_{ref}^{(\uparrow)}|\psi_{ref}^{(\uparrow)}>
-<\psi_{ref}^{(\downarrow)}|\psi_{ref}^{(\downarrow)}>\right).
\end{eqnarray}

Note, at $t=0,$ for spinor (\ref{1}), $\theta_{full}=\pi/2,$ $\phi_{full}=0.$
However, this is not the case for transmission and reflection. Namely, at $t=0$ we
have
\begin{eqnarray} \label{14}
\phi_{tr}^{(0)}=\arctan\left(\frac{\Im(<\psi_{tr}^{(\uparrow)}(x,0)|
\psi_{tr}^{(\downarrow)}(x,0)>)}
{\Re(<\psi_{tr}^{(\uparrow)}(x,0)|\psi_{tr}^{(\downarrow)}(x,0)>)}\right);
\end{eqnarray}
\begin{eqnarray} \label{15}
\theta_{tr}^{(0)}=\arccos\Big(<\psi_{tr}^{(\uparrow)}(x,0)|\psi_{tr}^{(\uparrow)}(x,0)>
\nonumber\\-<\psi_{tr}^{(\downarrow)}(x,0)|\psi_{tr}^{(\downarrow)}(x,0)>\Big);
\end{eqnarray}
\begin{eqnarray} \label{16}
\phi_{ref}^{(0)}=\arctan\left(\frac{\Im(<\psi_{ref}^{(\uparrow)}(x,0)|
\psi_{ref}^{(\downarrow)}(x,0)>)}
{\Re(<\psi_{ref}^{(\uparrow)}(x,0)|\psi_{ref}^{(\downarrow)}(x,0)>)}\right);
\end{eqnarray}
\begin{eqnarray} \label{17}
\theta_{ref}^{(0)}=\arccos\Big(<\psi_{ref}^{(\uparrow)}(x,0)|\psi_{ref}^{(\uparrow)}(x,0)>
\nonumber\\-<\psi_{ref}^{(\downarrow)}(x,0)|\psi_{ref}^{(\downarrow)}(x,0)>\Big).
\end{eqnarray}
Since the norms of $\psi_{tr}^{(\uparrow)}(x,t),$ $\psi_{tr}^{(\downarrow)}(x,t),$
$\psi_{ref}^{(\uparrow)}(x,t)$ and $\psi_{ref}^{(\downarrow)}(x,t)$ are constant,
$\theta_{tr}(t)=\theta_{tr}^{(0)}$ and $\theta_{ref}(t)=\theta_{ref}^{(0)}$ for any
value of $t$. For the (constant) $z$-components of spin we have
\begin{eqnarray} \label{18}
<\hat{S}_z>_{tr}(t)=\hbar\frac{T^{(\uparrow)}-T^{(\downarrow)}}{2T};\nonumber\\
<\hat{S}_z>_{ref}(t)=\hbar\frac{R^{(\uparrow)}-R^{(\downarrow)}}{2R}.
\end{eqnarray}

By our approach, the assertion made in \cite{But,Kob} that the magnetic field aligns
the particle's spin with the field is incorrect. We see that the influence of the
magnetic field on spin is twofold. Firstly, it causes the Larmor precession.
Secondly, it disturbs the balance between transmitted and reflected particles with
the up and down spins. As a result, transmitted and reflected particles possess a
nonzero $z$-component of the average spin, though for the case considered it is zero
for the whole ensemble of particles. However, we have to stress that the {\bf
ensemble} of transmitted (reflected) particles has the constant value of
$<\hat{S}_z>_{tr}(t)$ ($<\hat{S}_z>_{ref}(t)$) in the course of its (deterministic)
motion. This spin component must be constant, because the operator $\hat{S}_z$
commutes with Hamiltonian (\ref{200}) (we have here to bear in mind that the
individual probability fields for transmission and reflection, though superposed,
evolve independently each other).

To answer the question posed in \cite{Lea}, we have to pay reader's attention on the
following. As is seen from the above, of no importance is where the magnetic field
is localized: switching on the infinitesimal magnetic field, in any spatial region,
must change simultaneously the average spin of transmitted and reflected particles.

\section{Larmor precession under the infinitesimal magnetic field confined to the
barrier region}

So, the only effect caused by the magnetic field in this scattering process is the
Larmor precession of the average particle's spin. And our next step is to use these
"clocks" in order to define the time spent by particles in the barrier region. In
doing so, we have to take into account that the initial position of the
clock-pointer is not zero.

Following \cite{But,Kob} we will suppose that the applied magnetic field is
infinitesimal. Our first step is to find the derivations $d\phi_{tr}/dt$ and
$d\phi_{ref}/dt.$ For this purpose we will use the Ehrenfest equations for the
average spin of particles. One can show that
\[
\frac{d<\hat{S}_x>_{tr}}{dt}=-\hbar\omega_L \int_a^b
\Im[(\psi_{tr}^{(\uparrow)}(x,t))^*\psi_{tr}^{(\downarrow)}(x,t)]dx \]
\begin{eqnarray} \label{19}
\frac{d<\hat{S}_y>_{tr}}{dt}=\hbar\omega_L \int_a^b
\Re[(\psi_{tr}^{(\uparrow)}(x,t))^*\psi_{tr}^{(\downarrow)}(x,t)]dx
\end{eqnarray}
\[\frac{d<\hat{S}_x>_{ref}}{dt}=-\hbar\omega_L \int_a^{x_c}
\Im[(\psi_{ref}^{(\uparrow)}(x,t))^*\psi_{ref}^{(\downarrow)}(x,t)]dx\]
\[\frac{d<\hat{S}_y>_{ref}}{dt}=\hbar\omega_L \int_a^{x_c}
\Re[(\psi_{ref}^{(\uparrow)}(x,t))^*\psi_{ref}^{(\downarrow)}(x,t)]dx.\]

Note, \[\phi_{tr}= \arctan\left(\frac{<\hat{S}_y>_{tr}}{<\hat{S}_x>_{tr}}\right),
\phi_{ref}=\arctan\left(\frac{<\hat{S}_y>_{ref}}{<\hat{S}_x>_{ref}}\right)\] Hence
\[\frac{d \phi_{tr}}{dt}=\frac{\frac{d<\hat{S}_y>_{tr}}{dt} <\hat{S}_x>_{tr}
-\frac{d<\hat{S}_x>_{tr}}{dt}
<\hat{S}_y>_{tr}}{<\hat{S}_x>^2_{tr}+<\hat{S}_y>^2_{tr}}.\] The same expression
takes place for reflection.

However, in the case of the initial condition chosen and infinitesimal magnetic
field, when
\[\frac{|<\hat{S}_y>_{tr}|}{|<\hat{S}_x>_{tr}|}\ll 1\ooo and \ooo
\frac{|<\hat{S}_y>_{ref}|}{|<\hat{S}_x>_{ref}|}\ll 1,\] these expressions are
essentially simplified ---
\[\frac{d \phi_{tr}}{dt}=\frac{1}{<\hat{S}_x>_{tr}}\cdot \frac{d<\hat{S}_y>_{tr}}{dt};\]
\[\frac{d \phi_{ref}}{dt}=\frac{1}{<\hat{S}_x>_{ref}}\cdot \frac{d<\hat{S}_y>_{ref}}{dt}.\]
Considering Exps. (\ref{8})-(\ref{13}) and (\ref{19}), we obtain
\[\frac{d \phi_{tr}}{dt}=\omega_L \frac{\int_a^b
\Re[(\psi_{tr}^{(\uparrow)}(x,t))^*\psi_{tr}^{(\downarrow)}(x,t)]dx}
{\int_{-\infty}^\infty
\Re[(\psi_{tr}^{(\uparrow)}(x,t))^*\psi_{tr}^{(\downarrow)}(x,t)]dx};\]
\[\frac{d \phi_{ref}}{dt}=\omega_L \frac{\int_a^{x_c}
\Re[(\psi_{ref}^{(\uparrow)}(x,t))^*\psi_{ref}^{(\downarrow)}(x,t)]dx}
{\int_{-\infty}^{x_c}
\Re[(\psi_{ref}^{(\uparrow)}(x,t))^*\psi_{ref}^{(\downarrow)}(x,t)]dx}.\] Or, taking
into account that in the first order approximation, for the infinitesimal magnetic
field, when
$\psi_{tr}^{(\uparrow)}(x,t)=\psi_{tr}^{(\downarrow)}(x,t)=\psi_{tr}(x,t)$ and
$\psi_{ref}^{(\uparrow)}(x,t)= \psi_{ref}^{(\downarrow)}(x,t)=\psi_{ref}(x,t),$ we
have
\[\frac{d \phi_{tr}}{dt}\approx\omega_L \frac{\int_a^b
|\psi_{tr}(x,t)|^2dx} {\int_{-\infty}^\infty |\psi_{tr}(x,t))|^2 dx}>0;\]
\[\frac{d \phi_{ref}}{dt}\approx\omega_L \frac{\int_a^{x_c}
|\psi_{ref}(x,t)|^2dx}{\int_{-\infty}^{x_c} |\psi_{ref}(x,t)|^2dx}>0.\] As is seen,
the clocks chosen operate properly: their clock-pointers rotate in one direction.

As is supposed in our setting the problem, both at the initial and final moments of
time the ensemble of particles does not interact with the potential barrier and
magnetic field. In this case, without loss of exactness, the angles of rotation
($\Delta\phi_{tr}$ and $\Delta\phi_{ref}$) of spin under the magnetic field, in the
course of a completed scattering, can be written in the form,
\begin{eqnarray} \label{20}
\Delta\phi_{tr}=\frac{\omega_L}{T} \int_{-\infty}^\infty dt \int_a^b
dx|\psi_{tr}(x,t)|^2
\end{eqnarray}
\begin{eqnarray} \label{21}
\Delta\phi_{ref}=\frac{\omega_L}{R} \int_{-\infty}^\infty dt \int_a^{x_c}
dx|\psi_{ref}(x,t)|^2.
\end{eqnarray}
On the other hand, we have to bear in mind that the times spent, on the average, by
transmitted an reflected particles in the barrier region are finite. Let
$\tau^L_{tr}$ and $\tau^L_{ref}$ be the corresponding times for transmission and
reflection. Thus, $\Delta\phi_{tr}=\omega_L \tau^L_{tr}$ and
$\Delta\phi_{eef}=\omega_L \tau^L_{ref}.$ Comparing these expressions with
(\ref{20}) and (\ref{21}), respectively, we eventually obtain
\begin{eqnarray} \label{22}
\tau^L_{tr}=\frac{1}{T} \int_{-\infty}^\infty dt \int_a^b dx|\psi_{tr}(x,t)|^2
\end{eqnarray}
\begin{eqnarray} \label{23}
\tau^L_{ref}=\frac{1}{R} \int_{-\infty}^\infty dt \int_a^{x_c}
dx|\psi_{ref}(x,t)|^2.
\end{eqnarray}
These are just the searched-for definitions of the Larmor times for transmission and
reflection.

Unlike the group and dwell time concepts, the Larmor-time concept gives the basis to
verify our formalism. Indeed, let for the above setting the problem we have measured
the spin of transmitted particles. Let the azimuthal angle $\phi_{tr}(t)$ at $t\to
\infty$ be $\phi_{tr}^{(\infty)}$. By our approach, the final angle should be equal
to $\phi_{tr}^{(\infty)}=\phi_{tr}^{(0)}+\Delta\phi_{tr}$ (see Exps. (\ref{14}) and
(\ref{20})).

As is seen, there is the problem to distinguish the inputs $\phi_{tr}^{(0)}$ and
$\Delta\phi_{tr}.$ However, one can show, for example, that in the case of tunneling
a particle through an opaque rectangular potential barrier (just where the so-called
Hartman effect "appears') $|\phi_{tr}^{(0)}|\ll\Delta\phi_{ref}\ll\Delta\phi_{tr}.$
Thus, this case is most suitable and interesting for checking the above expressions
for the Larmor times.

\section{Connection between the Larmor and dwell times}

Note firstly that Exps. (\ref{22}) and (\ref{23}) for the Larmor times coincide by
form with those obtained in the known approaches (see, e.g.,
\cite{Lea,Nus,Hau,Le1,Ste,Aha}) as the dwell times. However, there are deep
differences between the known dwell-time concepts and Larmor times presented here.
Namely, 1) our Larmor times distinguish between transmitted an reflected particles;
2) they are exact, rather than asymptotic values; 3) they were derived for any
incident Gaussian-like wave packets, which may contain harmonics with the zero
momentum; 4) unlike the known times for transmission and reflection, they are real
and non-negative.

Let us now display the connection between Larmor time (\ref{22}), (\ref{23}) and the
dwell times derived in \cite{Ch2}. Let us consider the case of transmission. For
this purpose let us write down the wave function $\psi_{tr}(x,t)$ in the form
\begin{eqnarray} \label{24}
\psi_{tr}(x,t)=\frac{1}{\sqrt{2\pi}}\int_{-\infty}^{\infty} A(k)\psi_{tr}(x,k)e^{-i
E(k)t/\hbar}dk,
\end{eqnarray}
where $E(k)=\hbar^2k^2/2m$; $\psi_{tr}(x,k)$ is the stationary wave function for
transmission (see \cite{Ch1}); $A(k)$ is a real Gaussian-like function.

Let us transform the integral in (\ref{22}), \[I=\int_{-\infty}^\infty dt \int_a^b
dx|\psi_{tr}(x,t)|^2.\] Considering Exp. (\ref{24}) and integrating on $t$, we
obtain
\[I=\frac{\hbar}{\pi}\int_{-\infty}^{\infty} dk^\prime dk A(k^\prime) A(k)\int_a^b
dx \psi^*_{tr}(x,k^\prime) \psi_{tr}(x,k)\] \[\times\lim_{\Delta t\to\infty}
\frac{\sin[(E(k^\prime)-E(k))\Delta t/\hbar]}{E(k^\prime)-E(k)}.\] However,
\[\lim_{\Delta t\to\infty}
\frac{\sin[(E(k^\prime)-E(k))\Delta
t/\hbar]}{E(k^\prime)-E(k)}=\frac{\pi}{\hbar}\delta[(E(k^\prime)-E(k))/\hbar]\]
\[=\frac{\pi m}{\hbar^2 k}\left[\delta(k^\prime-k)-\delta(k^\prime+k)\right].\]
Hence
\begin{eqnarray} \label{25}
I=\frac{m}{\hbar}\int_{-\infty}^{\infty}dk \frac{A(k)}{k}\int_a^b dx
\Big[A(k)\psi^*_{tr}(x,k)\nonumber\\ - A(-k)\psi^*_{tr}(x,-k)\Big]\psi_{tr}(x,k).
\end{eqnarray}
The integrand in this expression is evident to be non-singular at $k=0$.

Let us consider the stationary case: $A(k)=\delta(k-k_0)$ (note, $k_0>0$). Then
\[I=\frac{m}{\hbar k_0}\int_a^b dx |\psi_{tr}(x,k_0)|^2.\] And, lastly, substituting
this expression in (\ref{22}), we obtain
\[
\tau^L_{tr}=\frac{m}{\hbar k_0 T}\int_a^b dx |\psi_{tr}(x,k_0)|^2. \] However, this
is just the dwell time for transmission (\cite{Ch1}).

In the same way one can show that in the stationary case the Larmor and dwell times
for reflection are equal too:
\[
\tau^L_{ref}=\frac{m}{\hbar k_0 R}\int_a^{x_c} dx |\psi_{ref}(x,k_0)|^2. \]

\section{Conclusion}

We have introduced the Larmor times to be valid for any symmetrical finite potential
barriers and Gaussian-like wave packets. The Larmor-time concept presented differs
from the known analogs. We show that in the case of the stationary tunneling problem
the Larmor times coincide with the dwell times obtained in our previous paper. The
formalism presented can be, in principle, verified experimentally.

For example, it is very important to measure the Larmor times for particles
tunneling through an opaque rectangular barrier. By our approach, in this case the
Larmor time for transmission should be very large, in comparison with the
predictions of other approaches.

\end{document}